\documentclass[10pt,letterpaper,twocolumn]{article} 

\usepackage{ol2}
\usepackage[draft]{hyperref}
\usepackage{amsmath}
\usepackage{bm,braket}

\begin{document}

\abovedisplayskip=5pt
\abovedisplayshortskip=5pt
\belowdisplayskip=5pt
\belowdisplayshortskip=5pt
\twocolumn[ 

\title{Mollow ``quintuplets'' from coherently-excited  quantum dots}


\author{Rong-Chun Ge$^1$, S. Weiler$^2$, A. Ulhaq$^{2,3}$, S.~M.~Ulrich$^2$, M. Jetter$^2$,
P. Michler$^2$, and\\ S. Hughes$^1$}
\address{
$^1$Department of Physics, Engineering Physics and Astronomy,
Queen's University, Kingston, Ontario, Canada K7L 3N6\\
$^2$Institut f\"ur Halbleiteroptik und Funktionelle Grenzfl\"achen, Allmandring 3, 70569 Stuttgart,Germany\\

$^3$Department of Physics School of Science and Engineering, Lahore University of
Management Sciences, Sector U, DHA, Lahore 54792 Pakistan
}

\begin{abstract}
Charge-neutral excitons in semiconductor quantum dots have a small
finite energy separation caused by the anisotropic exchange
splitting. Coherent excitation of neutral excitons will generally
excite both exciton components, unless the
excitation is parallel to one of the dipole axes. We present a
polaron master equation model to describe two-exciton pumping
using a coherent continuous wave pump field in the presence of a
realistic anisotropic exchange splitting. We predict a five-peak
incoherent spectrum, thus generalizing the Mollow triplet to
become a Mollow quintuplet. We experimentally confirm such
spectral quintuplets for  In(Ga)As quantum dots and obtain very good agreement with theory.
\end{abstract}


 ] 

\noindent
Quantum light-matter interaction in semiconductor nanostructures is a topic of considerable  interest. An understanding of the fundamental light-matter interactions is  important for future quantum devices, with applications ranging from  single photon emitters~\cite{Michler,Mor} entanglement-based sources~\cite{Ako06,Stevenson06}.  Due to  efficient semiconductor growth technology and natural scalability,  semiconductor quantum bits (QDs) offer an excellent opportunity to explore rich light-matter interaction in a solid state environment, and they   can function as   solid state quantum bits (qubits).
In the presence of high-field coherent optical pumping,
it is well known that the fluorescence spectrum of a driven two-level system develops a symmetric ``Mollow triplet'' structure~\cite{Mollow}, where the two outer sidebands are split by  the drive's Rabi frequency. 
While driven QDs, also referred to as  ``artifial atoms,'' do show behaviour that is similar to a driven two-level atom~\cite{Flagg.Muller:2009,Vamivakas.Atature:2009},
   due to the coupling of the exciton to a phonon reservoir~\cite{Axt:PRB2002,Foerstner.Weber:2003,Nazir:2008,imamoglu,nazir2,Roy:PRX11}, both the position and broadening of the Mollow triplets can change significantly~\cite{Roy:PRL11,Ata:arXiv}.  It is also well known that real QDs have many exciton states, and, e.g.,   charge-neutral excitons are split by a small anisotropic exchange energy. 
 
In this Letter we develop   a polaron master equation to include two neutral excitons and model the ensuing incoherent spectrum. We  generalize the well known Mollow triplet to a   {\em Mollow quintuplet} regime, which is caused by a sum of two separate Mollow triplets with a similar central   resonance.
A useful analytical expression for the incoherent spectrum  is given and compared to experiments, where we find very good qualitative agreement.
Figure 1 shows a schematic of the  excitation geometry.

\begin{figure}[b]
\centering
\includegraphics[width=.96\columnwidth]{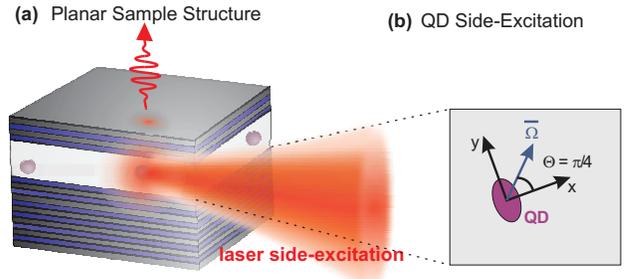}
\caption{Schematic of the QD under  continuous wave (cw) pumping with a field that is polarized along, e.g., $\theta=\frac{\pi}{4}$ in a weakly coupled planar cavity. This QD can be excited with electric dipole along the $x$ or $y$ axis.} 
\label{fig:QD}
\end{figure}

Neglecting QD zero-phonon-line (ZPL)  decay mechanisms, and
working in a rotating frame with respect to the  pump laser frequency $\omega_L$, the  two-exciton Hamiltonian is given by
$H=\sum_{j {\bf q}}\hbar\omega_{j{\bf q}}b^{\dagger}_{j{\bf q}}b_{j{\bf q}}-\sum_{j=1,2}\hbar\Delta_j\sigma_j^+\sigma_i^- + \sum_j\frac{\hbar\Omega_j}{2}(\sigma^+_j+\sigma^-_j)+\sum_{j,{\bf q}}\hbar\lambda_{j{\bf q}}\sigma_j^+\sigma_j^-\big(b^{\dagger}_{j{\bf q}}+b_{j{\bf q}}\big),$
where $\Delta_j\equiv\omega_L-\omega_{x_j}$ ($j=1,\, 2$) are the laser-exciton detunings, $b_{j{\bf q}}$ are the annihilation operators of the phonon reservoirs, $\sigma^+_j,\,\sigma^-_j$ are the Pauli operators of the $j$th exciton,  $\lambda_{j{\bf q}}$ represent the exciton-phonon coupling strength, and $\Omega_j$ are the Rabi frequencies of the continuous wave (cw) pump.

Adopting a polaron transformation $H'=e^{S}He^{-S}$, with $S=\sum_{j{\bf q}}\sigma_j^+\sigma_j^-\frac{\lambda_{j{\bf q}}}{\omega_{j{\bf q}}}\big(b_{j{\bf q}}^{\dagger}-b_{j{\bf q}}\big)$, the exciton-phonon coupling can be taken into account nonperturbatively~\cite{imamoglu,nazir2}.
To second-order in the polaron-transformed  exciton-phonon interaction, the polaron master equation can be written as
$
\frac{\partial \rho(t)}{\partial t}=\frac{-i}{\hbar}[H'_{\rm sys},\rho(t)]+\sum_j\frac{\gamma_j}{2}\mathcal{L}[\sigma^-_j]+
\sum_j\frac{\gamma'_j}{2}\mathcal{L}[\sigma^{11}_j]
-\frac{1}{\hbar^2}\int_0^{\infty}d\tau(\left[H_I, H_I(-\tau)\rho(t)\right]
+{\rm H.c.}) \, ,
$
where
the upper time limit of the time integral, $t\rightarrow\infty$, since
the relaxation time of acoustic phonon bath is very fast (a few ps). The polaron-modified system Hamiltonian,
$H_{\rm sys}'=-\sum_{j=1,2}\hbar(\Delta_j+\Delta_{p_j})\sigma_j^+\sigma_j^- + \sum_j\frac{\hbar\Omega_j'}{2}(\sigma^+_j+\sigma^-_j)$, where
the renormalized Rabi frequency is given by ${\Omega'}_{j}=\Omega_j\langle B_j\rangle$, with  $\braket{B_j}={\rm exp} \left [\frac12\int_0^{\infty}d\omega\frac{J(\omega)}{\omega^2}\coth(\frac{\hbar\omega}{2k_bT})\right ]$
which depends on the temperature of the phonon bath.
We chose a phonon spectral function that accounts for exciton-LA interactions from
 the deformation potential,  $J_j(\omega)=\alpha_{p_j}\omega^3\exp (-{\omega^2}/{2\omega^2_{b_j}} )$~\cite{Ulh10}, where $\omega_{b_j}$ is the phonon cutoff frequency and $\alpha_{p_j}$ characterizes the strength of the exciton-phonon interaction.
 We  also include Lindblad scattering terms, $\mathcal{L}[D]=(D\rho\, D^{\dagger}-D^{\dagger}D\rho)+{\rm H.c.}$, which  describe ZPL radiative decay and ZPL pure dephasing~\cite{BorriPRL:2001}.
For convenience, we will  include the polaron shift $\Delta_{jp}=\int_0^{\infty}d\omega\frac{J_j(\omega)}{\omega}$ into $\Delta_j$ below.

 For Rabi fields much less than $\omega_{b_j}$,
one can derive an  {\em effective phonon master equation}~\cite{Roy:PRX11,Ata:arXiv},
\begin{align}
&\frac{\partial \rho}{\partial t}=\frac{-i}{\hbar}\left [ H'_{\rm sys}, \rho\right ] + \sum_j\frac{\gamma_j}{2}\mathcal{L}[\sigma^-_j]+\sum_j\frac{\gamma'_j}{2}\mathcal{L}
[\sigma^{+}_j\sigma^{-}_j]\nonumber\\
&+
\sum_j\frac{\Gamma^{\sigma^+_j}_{\rm ph}}{2}\mathcal{L}[\sigma^+_j]+\sum_j\frac{\Gamma^{\sigma^-_j}_{\rm ph}}
{2}\mathcal{L}[\sigma^-_j] -\sum_j\Gamma^{\rm cd}_{{\rm ph},j}
{\cal S},
\label{eq:ME}
\end{align}
where ${\cal S}=(\sigma^+_j\rho\sigma^+_j +
\sigma^-_j\rho\sigma^-_i)$. This model is  applicable to two
separate QDs or two excitons from the same dot, and we will focus
on the later here to model realistic single QDs with driven
neutral excitons. The phonon-induced scattering rates  are derived
as follows~\cite{Roy:PRX11,Ata:arXiv}:
$\Gamma^{\sigma^{\pm}_j}_{\rm ph}=\frac{{\Omega'}_{j}^2}{2}{\rm
Re}\left [\int^{\infty}_0d\tau e^{\pm
i\Delta_j\tau}\left(e^{\phi(\tau)}-1\right)\right]$ and
$\Gamma^{\rm cd}_{{\rm ph},j}=\frac{{\Omega}_{j}^{'2}}{2} {\rm
Re}\left
[\int^{\infty}_0d\tau\cos(\Delta_j\tau)\left(1-e^{-\phi(\tau)}\right)\right]$,
where
$\phi(\tau)=\int_0^{\infty}d\omega\frac{J(\omega)}{\omega^2}\left[\coth(\frac{\beta\hbar\omega}{2})
\cos(\omega\tau)-i\sin(\omega\tau)\right]$. For the one exciton
limit, the above model has been used to  describe the spectra for
driven In(Ga)As QDs~\cite{Ata:arXiv}, where phonon-mediated
processes are also found to play a significant role. Consistent with these
 experiments, we will use 
$\omega_{b_1}=\omega_{b_2}=1$~meV and $\alpha_{p_1} = \alpha_{p_2}
= 0.15/(2\pi)^2\, {\rm ps}^{2}$.

The emission from the QD is experimentally detected through a planar cavity (which spatially separates the pump field from the emitted spectrum), and the incoherent spectrum of the system is given by~\cite{Ata:arXiv} $S ({\bf r},\omega) \equiv F({\bf r})S(\omega)$. Here $F({\bf r})$ is a geometrical factor, and $S(\omega)$ can be calculated by the quantum regression theorem.
From our two-exciton polaron master equation,
the  analytical spectrum is calculated to be
\begin{align}
S(\omega)&=\sum_{j}\beta_j\,\Bigg\{{\rm Re}\left[\frac{ih_i(0)C_j(\omega)D_j(\omega)-f_j(0)D_j(\omega)}
{(D_j(\omega)+i2\Delta_j)D_j(\omega)-E_j(\omega)^2}\right]\nonumber\\
&-{\rm Re}\left[\frac{E_j(\omega)\left[g_j(0)+
ih_j(0)C_j(\omega)\right]}
{(D_j(\omega)+i2\Delta_j)D_j(\omega)-E_j(\omega)^2}\right]\Bigg\},
\end{align}
 where  $C_j(\omega)={\Omega'_{j}}/{[2(i\delta\omega-\gamma_{\rm pop}^j)]}$, $D_j(\omega)=i\delta\omega-\gamma_{\rm pol}^j-i\Delta_j+{{\Omega'}_{j}^2}/{[2(i\delta\omega-\gamma_{\rm pop}^j)]}$, and $E_j(\omega)=\Gamma^{\rm cd}_{{\rm ph},j}+\Omega'_{1}C_j(\omega)$.
The  steady-state functions are  $f_j(0)\equiv\langle\delta\sigma^+_j\delta\sigma^-_j\rangle_{\rm ss}=\frac12\left[1+\langle\sigma^z_j\rangle_{\rm ss}-2\langle\sigma^-_j\rangle_{\rm ss}\langle\sigma^+_j\rangle_{\rm ss}\right] $, $g_j(0)\equiv\langle\delta\sigma^+_j\delta\sigma^+_j\rangle_{\rm ss}=-\langle\sigma^+_j\rangle_{\rm ss}^2$, and $h_j(0)\equiv\langle\delta\sigma^+_j\delta\sigma^z_j\rangle_{\rm ss}=-\langle\sigma^+_j\rangle_{\rm ss}\left[1+\langle\sigma^z_j\rangle_{\rm ss}\right]$, where
the steady-state inversion and polarization are given by
$
\langle\sigma^z_j\rangle_{\rm ss} ={-\big(\gamma_{\rm pop}^j-2\Gamma^{\sigma^+_j}_{\rm ph}\big)}/\big [{\gamma_{\rm pop}+\frac{{\Omega'}_{j}^2(\gamma_{\rm cd}+\gamma_{\rm pol})}
{\gamma_{\rm pol}^2+\Delta_j^2-\gamma_{\rm cd}^2}}\big]$
and
$
\langle\sigma^-_j\rangle_{\rm ss} =\frac{i{\Omega'}_{j}\left(\gamma_{\rm pol}+i\Delta_j+\gamma_{\rm cd}\right)}
{2\left(\gamma_{\rm pol}^2+\Delta_j^2-\gamma_{\rm cd}^2\right)}\langle\sigma^z_j\rangle_{\rm ss}.$
  The polarization and population decay rates are defined through $2\gamma_{\rm pol}^j={\gamma_j+\gamma'_j+\Gamma^{\sigma^+_j}_{\rm ph}+\Gamma^{\sigma^-_j}_{\rm ph}}$ and $\gamma_{\rm pop}^j=\gamma_j+\Gamma^{\sigma^+_j}_{\rm ph}+\Gamma^{\sigma^-_j}_{\rm ph}$.
Finally,
  $\alpha_j$ are   phenomenal scaling terms that adjust the strength of the exciton emission through  cavity decay.

\begin{figure}[t]
\centering\includegraphics[width=1\columnwidth]{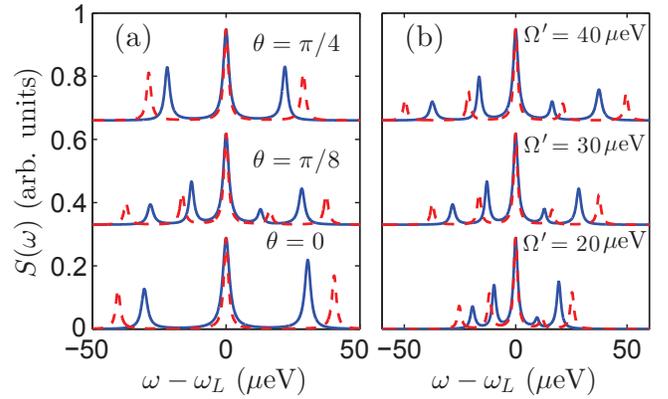}
\caption{(Color online) Incoherent spectrum $S(\omega)$ with (blue, solid) and without (red, dashed) phonon scattering. The  Rabi frequencies are $\Omega'_1=\Omega'\cos{\theta}$ and $\Omega'_2=\Omega'\sin{\theta}$, and $\gamma_{1,2}=\gamma'_{1,2}=1$~$\mu$eV,  $\delta_{12}=-12$~$\mu$eV, and  $\alpha_{1,2}=1$. (a) $\Omega'=30~\mu$eV, with different $\theta$. (b) $\theta=\pi/8$, with different $\Omega'$.}
\label{fig:10002}
\end{figure}

 Due to Coulomb interactions and orientational inhomogeneities of the QDs, 
there are two neutral excitons polarized along the $x$ and $y$ axes, respectively [c.f.~Fig.~\ref{fig:QD}].
 Thus for various pump polarization angles, $\theta$, there will be different  Rabi frequencies, $\Omega'_1=\Omega_1'\cos{\theta}$ and $\Omega'_2=\Omega'\sin{\theta}$, for the two excitons. Here $\Omega_{i}'=\braket{B}\Omega_i^{0}$ is the maximum Rabi frequency. For clarify, we will first assume that
$\Omega_{1}'=\Omega_{2}' = \braket{B}\Omega^{0}\equiv\Omega'$, though in general these will differ because the dipole moments will in general be different.
Figure \ref{fig:10002}(a) shows an example calculation of the
incoherent spectrum for various $\theta$, using  $\Delta_1=-\Delta_2=-6$~$\mu$eV.
We assume a phonon bath temperature of $T=6~$K, and use
parameters for In(Ga)As QDs~\cite{steffi:OE12}.
The main  parameters are given in figure caption.
 The red dashed lines shown in the
same figure are the results obtained without phonon scattering
(apart from ZPL decay), which show that phonon scattering plays a
qualitatively important role; in particular, we see
phonon-mediated spectral broadening and phonon renormalization
induced by phonon scattering. For $\theta=0,$ only one exciton is
excited, and the spectrum shows  the expected Mollow triplet; note that the spectrum  is asymmetric since the pump field is off resonance.
As $\theta$ increases, then the other orthogonal exciton
is gradually excited, as can be clearly seen  at
$\theta=\pi/8$---where the Mollow triplet evolves into a {\em
Mollow quintuplet}. At $\theta=\pi/4$, the quintuplets merge into
a triplet again if the two excitons share  symmetric parameters.
Figure \ref{fig:10002}(b) shows the spectrum at
$\theta=\frac{\pi}{8}$ for various Rabi fields
($\Omega'_0=20,\,30,\,40$~$\mu$eV from bottom to top), which show
how the spectral quintuplets evolves with  pump strength. From the
analytical spectrum, it can be found that there will be six peaks
for the incoherent spectrum in general. However the central peak
(for both excitons) will always lie at the laser pump frequency
and show one peak, so there are actually five peaks. 


Next we turn our attention to  experiments.
The planar sample under investigation is grown by metal-organic
vapor epitaxy. Self-assembled In(Ga)As QDs are embedded in a GaAs
$\lambda$-cavity, sandwiched between 29 (4) periods of
$\lambda/4$-thick AlAs/GaAs layers as the bottom (top) distributed
Bragg reflectors. For the investigations, the sample is kept in a
Helium flow cryostat providing high temperature stability $T = 6
\pm 0.5$~K. Laser stray-light suppression is achieved by use of an
orthogonal geometry between QD excitation and detection in
combination with polarization suppression and spatial filtering
via a pinhole. Resonant QD excitation is achieved by a narrow-band
($\approx 500$~kHz) continuous-wave (cw) Ti:Sapphire ring laser.
For high-resolution spectroscopy (HRPL) of micro-photoluminescence
($\mu$-PL) we employ a scanning Fabry-P\'{e}rot interferometer
with $\Delta E^{\rm HRPL}_{\rm res}< 1\,\mu$eV as described
earlier \cite{Ata:arXiv,steffi:OE12}.

 \begin{figure}[t]
\centering\includegraphics[width=1\columnwidth]{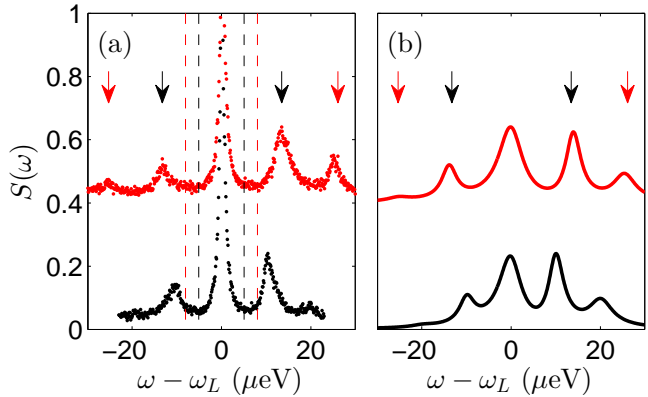}
\caption{(Color online) Experimental data for a driven In(Ga)As QD
and corresponding theoretical simulations, both as a  phonon bath
temperature of 6\,K. (a) Experimental results  with pump powers
$50~\mu$W (dark, lower) and $100~\mu$W (red, upper). (b) Theoretical
fits, with the following parameters: $\gamma_1=0.55$~$\mu$eV,
$\gamma_2=0.85$~$\mu$eV, $\gamma'_1=5.93$~$\mu$eV,
$\gamma'_2=7.13$~$\mu$eV, $\delta_{12}=10$~$\mu$eV. The Rabi
frequencies are $\Omega'_1={13.66}/{\sqrt{2}}$~$\mu$eV,
$\Omega'_2={21.96}/{\sqrt{2}}$~$\mu$eV (dark, lower), and
$\Omega'_1=13.66$~$\mu$eV, $\Omega'_2=21.96$~$\mu$eV (red, upper);
we also choose $\alpha_1=1$ and $\alpha_2=0.4$.} \label{fig:10001}
\end{figure}

Figures \ref{fig:10001}(a)-(b) show the experimental results and
theoretical simulations, respectively, for two different pump
fields.  Due to the fact the experimental central peak includes
contributions from both coherent and incoherent,  we focus on
reproducing the experimental data around the Mollow sidebands.
Figure~\ref{fig:10001}(a) is the experimental results  under pump
strength $50~\mu$W (dark, lower) and $100~\mu$W (red, upper). The
dark (inner) arrows and red (outer) arrows show the inner sideband
and outer sideband of the quintuplet, respectively.
Figure~\ref{fig:10001}(b) shows the theoretical simulations with
fitting parameters given in the caption, which are  consistent with earlier experiments \cite{Ata:arXiv,steffi:OE12}. 
We obtain a very good qualitative agreement. Moreover, we have observed these spectral
quintuplets from many of our In(Ga)As QDs. 

In summary, we have introduced a theory to describe resonance
fluorescence of coherently excited QD neutral excitons and predict
the emergence of a spectral quintuplet. The theory  uses a polaron master
equation from which an analytical spectrum is derived and presented. Using  data for
In(Ga)As QDs, we  obtain excellent agreement between theory
and experiment. \vspace{0.2cm}

This work was supported by the National Sciences and Engineering Research Council of Canada,
 the Carl-Zeiss-Stiftung, and the DFG (contract No. 500/23-1 MI.)



\begin{thebibliography}{99}

\bibitem{Michler}
P. Michler, A. Kiraz, C. Becher, W. V. Schoenfeld, P. M. Petroff, Lidong Zhang, E. Hu, A. Imamo\ifmmode \breve{g}\else \u{g}\fi{}lu, Science 
{\bf 290}, 2282 (2000).

\bibitem{Mor}
E. Moreau, I. Robert, J. M. G\'erard, I. Abram, L. Manin, and V. Thierry-Mieg, Appl. Phys. Lett. {\bf 79}, 2865 (2001).



\bibitem{Ako06}
N. Akopian, N. H. Lindner, E. Poem, Y. Berlatzky, J. Avron, and D. Gershoni, Phys. Rev. Lett. {\bf 96}, 130501 (2006).

\bibitem{Stevenson06}
R. M. Stevenson, 
R. J. Young, P. Atkinson, K. Cooper, D. A. Ritchie and A. J. Shields,
Nature {\bf 439}, 179 (2006).

\bibitem{Mollow}
B. R. Mollow, Phys. Rev. {\bf 188}, 1969 (1969).



\bibitem{Flagg.Muller:2009} E. B. Flagg,  A. Muller,  J. W. Robertson,  S. Founta,  D. G. Deppe,  M. Xiao,  W. Ma,  G. J. Salamo, and  C. K. Shih,
Nat. Phys. {\bf 5} 203 (2009).

\bibitem{Vamivakas.Atature:2009} A. Nick Vamivakas,  Yong Zhao,  Chao-Yang Lu, and  Mete Atat\"ure,
Nat. Physics {\bf 5}, 198 (2009).


\bibitem{Axt:PRB2002}
A. Vagov, V. M. Axt, and T. Kuhn,
Phys. Rev. B {\bf 66}, 165312 (2002).

\bibitem{Foerstner.Weber:2003}
C. F\"{o}rstner, C. Weber, J. Danckwerts, and A. Knorr,
Phys. Rev. Lett. {\bf 91}, 127401 (2003).

\bibitem{Nazir:2008}
A. Nazir,
Phys. Rev. B {\bf 78}, 153309, (2008).

\bibitem{imamoglu}I. Wilson-Rae,  and A. Imamo\ifmmode \breve{g}\else \u{g}\fi{}lu,
 Phys. Rev. B {\bf65}, 235311 (2002).

\bibitem{nazir2}D. P. S. McCutcheon, and A. Nazir,
New J. Phys. {\bf 12}, 113042 (2010).

\bibitem{Roy:PRX11}
C. Roy, and S. Hughes, Phys. Rev. X {\bf 1}, 021009 (2011).


\bibitem{Ata:arXiv} A. Ulhaq, S. Weiler, C. Roy, S. M. Ulrich, M. Jetter, S. Hughes, and P. Michler,  Opt. Express, in press.




\bibitem{Roy:PRL11}
C. Roy and S. Hughes, Phys. Rev. Lett. {\bf 106}, 247403 (2011).


\bibitem{Ulh10}
A. Ulhaq, S. Ates, S. Weiler, S. M. Ulrich, S. Reitzenstein, A. F\"offler, S. H\"ofling, L. Worschech, A. Forchel,
and P. Michler, Phys. Rev. B {\bf 82}, 045307 (2010).


\bibitem{BorriPRL:2001} P. Borri, W. Langbein, S. Schneider, U. Woggon, R. L. Sellin, D.
Ouyang, and D. Bimberg,
Phys. Rev. Lett. {\bf 87}, 157401 (2001).


\bibitem{steffi:OE12}
S. Weiler, A. Ulhaq, S. M. Ulrich, D. Richter, M. Jetter, P. Michler, C. Roy, and S. Hughes,
  Phys. Rev. B {\bf 86}, 241304(R) (2012).






\end{thebibliography}
\end{document}